\documentclass[apjl]{emulateapj}
\slugcomment{Accepted for ApJ Letters }

\shorttitle{A first constraint on the thick disk scale-length}
\shortauthors{BENSBY ET AL.}

\newcommand\teff{T_{\rm eff}}
\newcommand\kms{\rm\,km\,s^{-1}}

\newcommand\ulsr{U_{\rm LSR}}
\newcommand\vlsr{V_{\rm LSR}}
\newcommand\wlsr{W_{\rm LSR}}

\begin{document}

\title{
A first constraint on the thick disk scale-length:\\
Differential radial abundances in K giants at\\ 
Galactocentric radii 4, 8, and 12 kpc\altaffilmark{1}
}
\author{T. Bensby\altaffilmark{2}, A. Alves-Brito\altaffilmark{3}, M.S. Oey\altaffilmark{4}, D. Yong\altaffilmark{5}, \and J. Mel\'endez\altaffilmark{6}}
\affil{
\altaffilmark{2}Lund Observatory, Department of Astronomy and Theoretical Physics, Box 43, SE-221\,00 Lund, Sweden\\
\altaffilmark{3}Departamento de Astronom\'ia y Astrof\'isica, Pontificia Universidad
Cat\'olica de Chile, Santiago, Chile\\
\altaffilmark{4}Department of Astronomy, University of Michigan, Ann Arbor, 
MI 48109-1042, USA\\
\altaffilmark{5}Research School of Astronomy and Astrophysics, Australian National  University,
Weston, ACT 2611, Australia\\
\altaffilmark{6}Departamento de Astronomia do IAG/USP, Universidade de S\~ao Paulo,
Rua do Mat\~ao 1226, S\~ao Paulo, 05508-900, SP, Brasil
}

\altaffiltext{1}{
This paper includes data gathered with the 6.5 meter Magellan
Telescopes located at the Las Campanas Observatory, Chile
}

\begin{abstract}
Based on high-resolution spectra obtained with the MIKE spectrograph
on the Magellan telescopes we present detailed elemental abundances
for 20 red giant stars in the outer Galactic disk, located at Galactocentric
distances between 9 and 13\,kpc. The outer disk sample is complemented with 
samples of red giants from the inner Galactic disk and the solar neighbourhood, 
analysed using identical methods. For Galactocentric distances beyond 10\,kpc, 
we only find chemical patterns associated with the local thin disk, even for stars far 
above the Galactic plane.  Our results show that the relative densities of the thick and thin disks are dramatically different from the solar neighbourhood, and
we therefore suggest that the radial scale length of the thick disk is
much shorter than that of the thin disk. We make
a first estimate of the thick disk scale-length of 
$L_{thick}=2.0$\,kpc, assuming $L_{thin}=3.8$\,kpc for the thin disk.
We suggest that radial migration may explain the lack of 
radial age, metallicity, and abundance gradients in the thick disk, 
possibly also explaining the link between the thick disk 
and the metal-poor bulge.
\end{abstract}

\keywords{
Galaxy: formation ---
Galaxy: evolution ---
Galaxy: disk ---
Galaxy: stellar content ---
Galaxy: abundances ---
stars: abundances
}

\section{Introduction}

Thick disks in external galaxies were discovered by 
\cite{burstein1979} when the vertical light 
profiles of a few edge-on spiral galaxies could not be fitted by single
exponentials. Similarly, the Galactic thick disk was first detected when 
star count data towards the South 
Galactic Pole could not be fitted with one power law but two were needed
\citep{gilmore1983}. As there is no 
a priori law that says that the vertical star counts in spiral galaxies 
must fit single power laws this finding was necessary but not sufficient 
to define the thin and thick disks as a unique entities. The duality of 
the Galactic disk has been further seen in its kinematic and chemical 
properties: the thick disk lags the LSR by $\approx 40-50\,\kms$; the 
thick disk is more metal-poor than the thin disk \citep[e.g.,][]{gilmore1995,wyse1995};
the thick disk is older than the thin disk 
\citep[e.g.,][]{fuhrmann2008,bensby2007letter2}; and the thick disk is
$\alpha$-enhanced, at a given metallicity, with respect to the thin disk 
\citep[e.g.,][]{fuhrmann2008,bensby2003,bensby2004,bensby2005,reddy2006}.

\begin{figure*}
\centering
\resizebox{0.9\hsize}{!}{
\includegraphics[bb=18 155 592 710,clip]{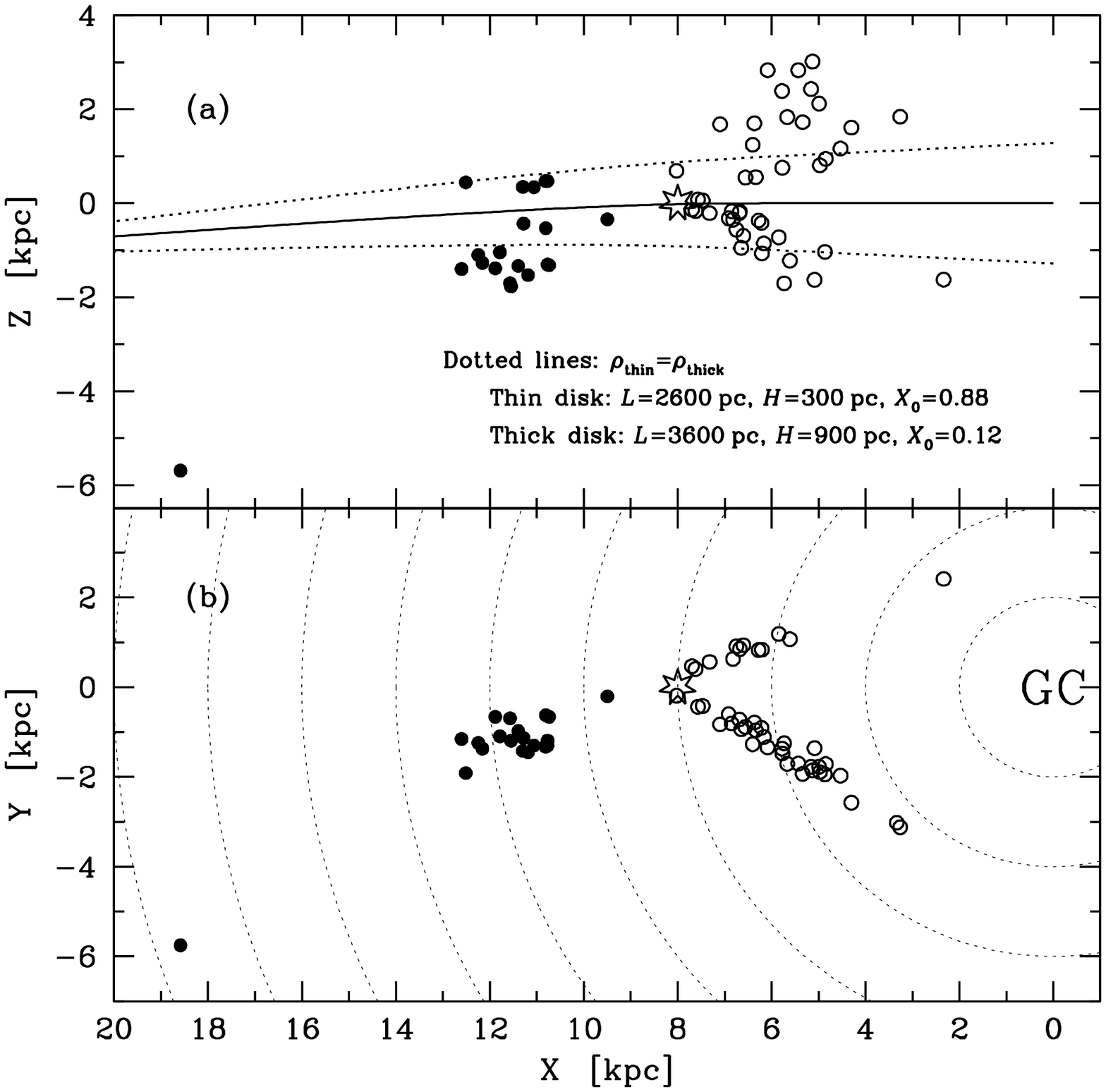}
\includegraphics[bb=0 155 592 710,clip]{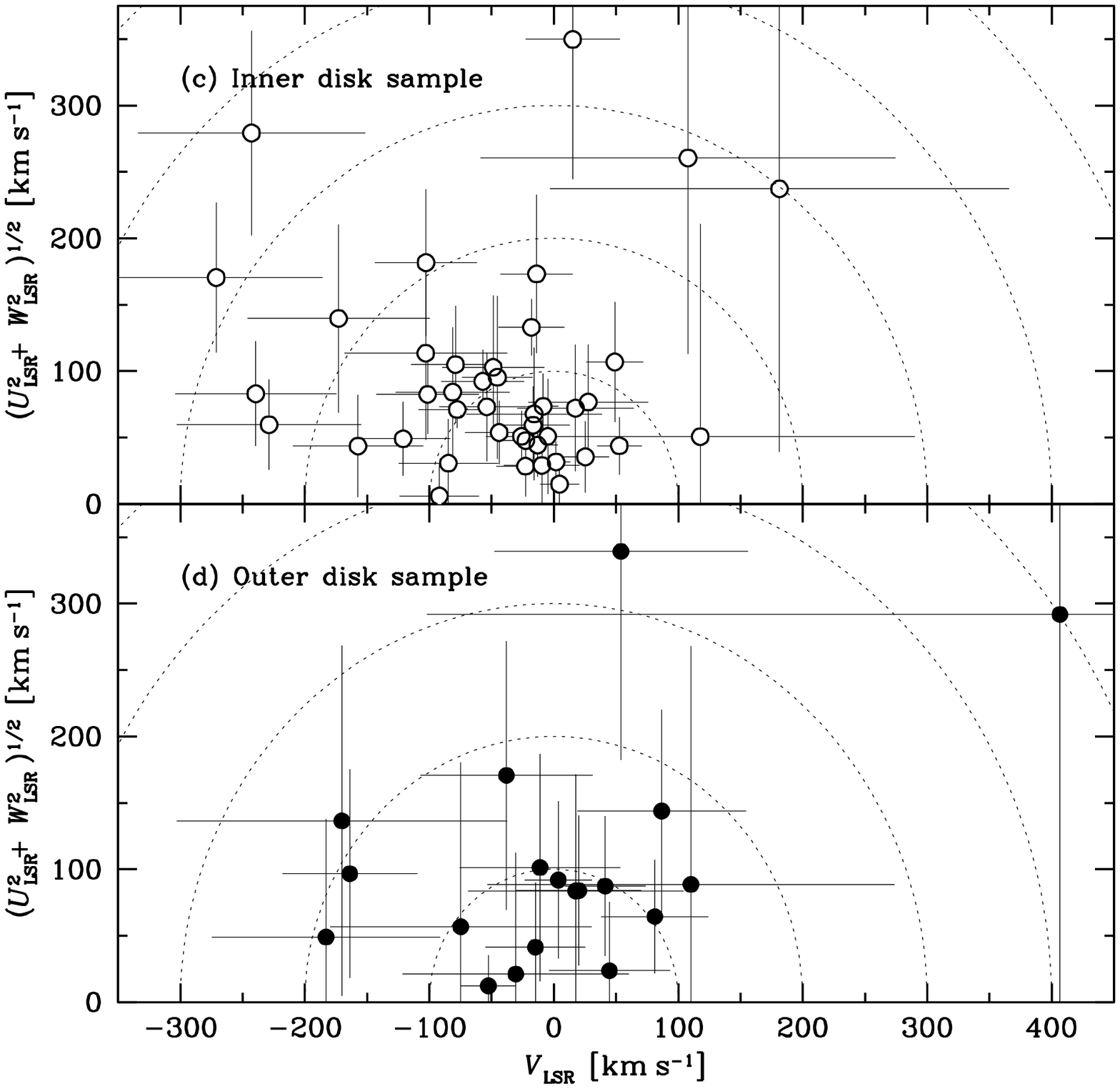}}
\caption{(a) and (b) show the locations of the stars in Galactic $X$, $Y$, 
and $Z$ coordinates. Outer disk stars are marked by filled circles
and the inner disk stars from \cite{bensby2010letter} by open circles.
Dotted lines in (a) represent the
distances above and below the plane where the thin and thick disk 
stellar densities are equal, given the scale-lengths, scale-heights, 
and normalisations for the thin and thick disks given by \cite{juric2008}.
The warp of the disk as given by \cite{momany2006} has been included. 
(c) and (d) show Toomre diagrams
for the inner and outer disk samples, respectively.
\label{fig:glonglat} 
}
\end{figure*}

The above studies are based on stellar samples within
the solar cylinder, i.e. at Galactocentric distances ($R_{\rm G}$)
around 8\,kpc. The inner and outer regions of the Galactic disk are 
far less studied. Actually, the inner disk is one of the least studied
regions of the Milky Way due to the high interstellar extinction and
contamination by background bulge stars. Apart from a few studies
of bright hot OB stars \citep[e.g.,][]{daflon2004} and Cepheids
\citep[e.g.,][]{luck2006}, that trace the young disk
stellar population, the only available data on the
abundance structure of the inner Galactic disk is from 
\cite{bensby2010letter} who studied 44 red giants located 3 to 7\,kpc 
from the Galactic centre, and found evidence for a similar
duality as seen in the solar neighbourhood.

The outer disk is comparatively well studied, especially using red giants in open clusters
\citep[e.g.,][and references therein]{yong2005,jacobson2011}, and to very large 
$R_{\rm G}$ indeed ($>20$\,kpc, \citealt{carraro2007}).
Also OB stars  \citep[e.g.,][]{daflon2004,daflon2004outer}, and Cepheids 
\citep[e.g.,][]{andrievsky2004,yong2006} have been observed in the outer disk.
\cite{carney2005} observed 
three outer disk field red giants, which turned out to have
abundance ratios similar to those in outer disk open clusters. 
These studies show an abundance gradient which is very steep
inside $R_{\rm G}\approx 10$\,kpc. For distances greater than $R_{\rm G}>10$\,kpc
it is less step, or 
possibly even flat, converging on metallicity around $\rm [Fe/H]\approx -0.3$
(see also \citealt{twarog1997} and the compilation by \citealt{cescutti2007}). 
Open clusters, OB stars, and Cepheids all
trace the young stellar population of the disk, and it is therefore unclear wether
the outer disk shows a similar duality
as observed in the solar neighbourhood and in the inner disk. 

This letter extends the study on red giants in the inner disk by
\cite{bensby2010letter} to include red giants in the outer disk.
Detailed elemental abundances are presented for 20 red giant stars,
located at $R_{\rm G}$ between 9 and 13\,kpc, and 0.5 to 2\,kpc from the Galactic
plane. 
They have been analysed using the exact same methods as used in the study 
of inner disk giants by \cite{bensby2010letter}  and in the study of  red 
giants in the Bulge and the nearby thin and thick 
disks by \cite{alvesbrito2010}. This allows for a truly differential comparison,
free from systematic offsets and uncertainties between the different 
stellar samples.

\section{Sample selection, observations, and analysis}

\subsection{Methods}

Twenty red giants in the outer disk, selected from the 2MASS
catalogue, were observed with the MIKE spectrograph \citep{bernstein2003}
on the Magellan telescopes
in Nov. 2007. Sample selection, instrumental setup, data reduction, abundance analysis,
determination of  distances, space velocities, and orbital parameters
were done in exactly the same way as for the 44 inner disk red giants
in \cite{bensby2010letter},
where we direct the reader for the details. 
Table~\ref{tab:parameters} gives
the stellar parameters, kinematics, 
and elemental abundances for the 20 outer disk giants {\it and}
the 44 inner disk giants from \cite{bensby2010letter}.

\begin{figure*}
\centering
\resizebox{\hsize}{!}{
\includegraphics[bb= 0 160 445 620,clip]{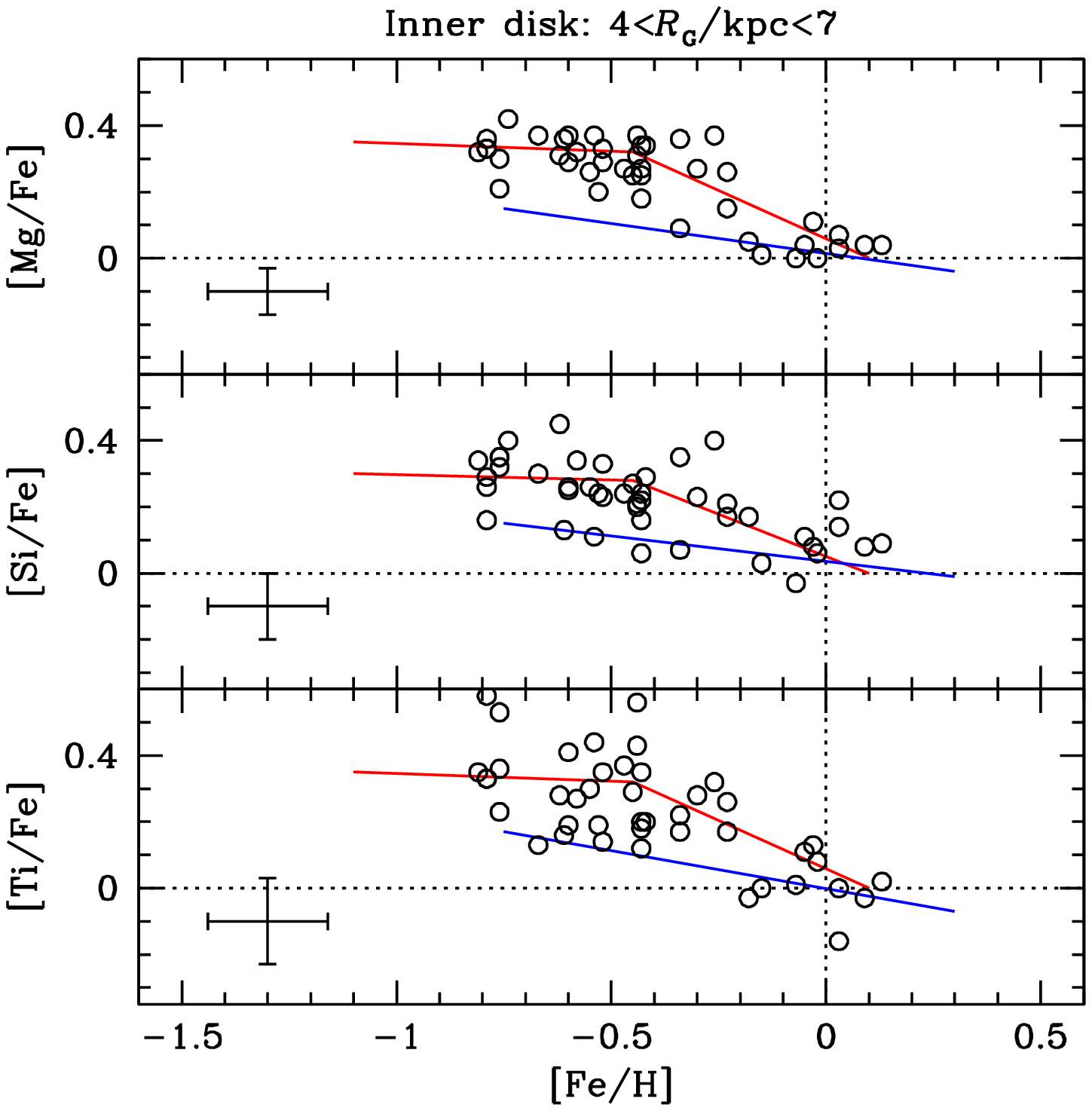}
\includegraphics[bb=75 160 445 620,clip]{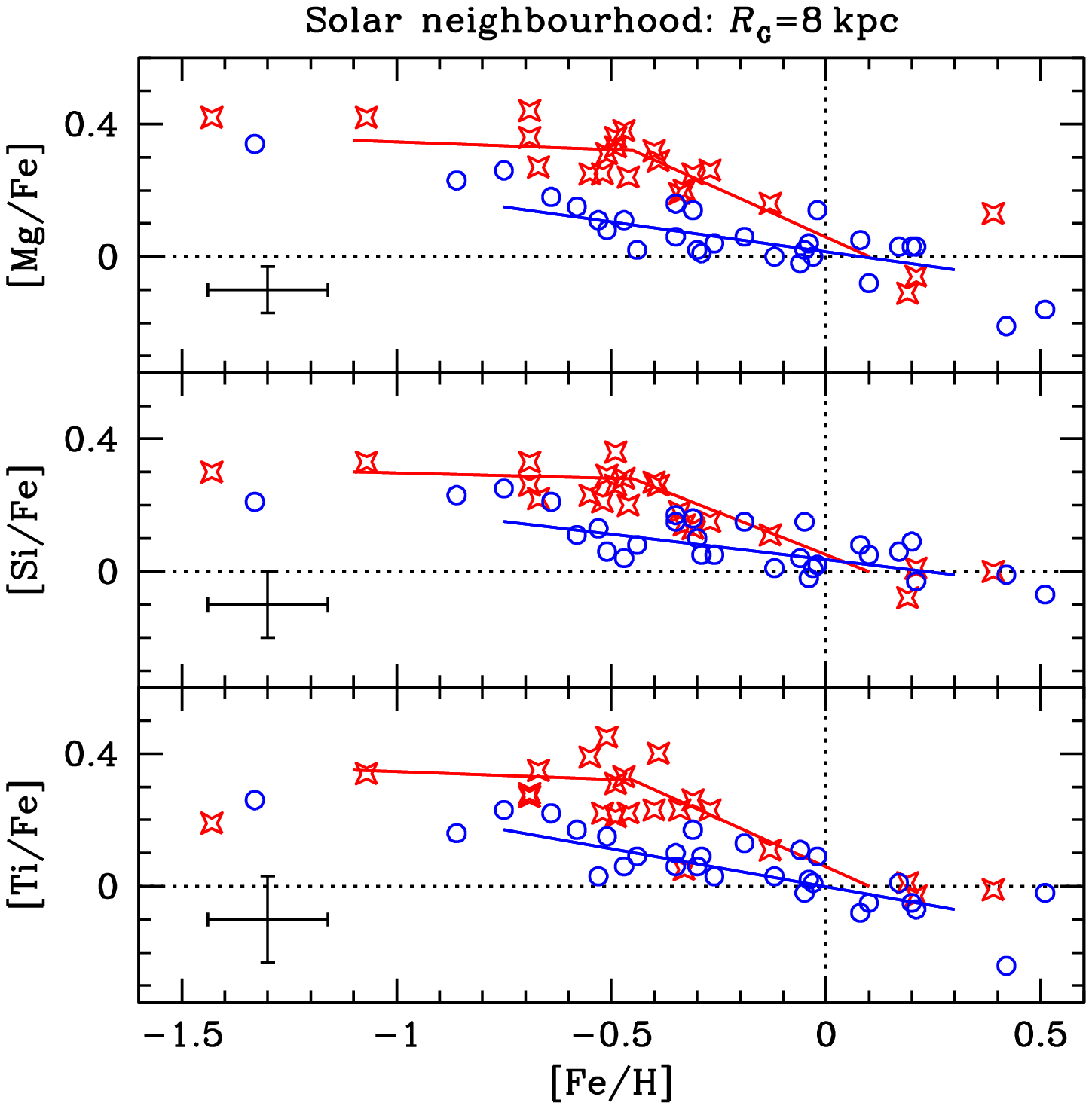}
\includegraphics[bb=75 160 465 620,clip]{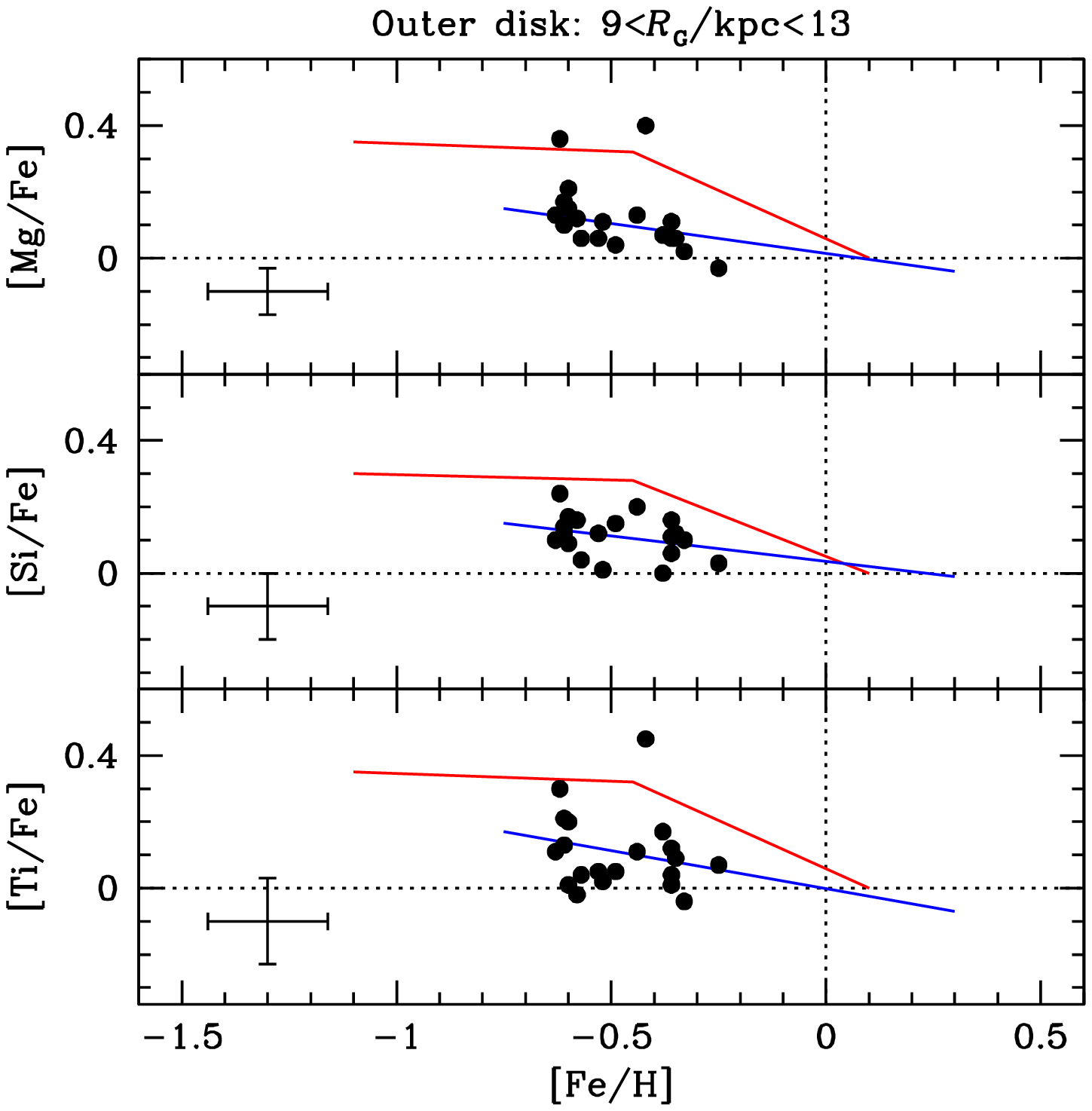}}
\caption{
Abundance trends for the $\alpha$-elements Mg, Si, and Ti. 
The left panel shows the 44 inner disk red giants from 
\cite{bensby2010letter}, the centre panel shows the solar neighbourhood
thin and thick disk stars 
(blue circles and red stars, respectively) by \cite{alvesbrito2010}. 
The right panel shows the 20 new 
outer disk red giants.  The red and blue lines in the abundance
plots are fiducial lines based on the solar neighbourhood abundance 
trends. 
\label{fig:haltplottar} 
}
\end{figure*}
\begin{figure}
\centering
\resizebox{0.9\hsize}{!}{
\includegraphics[bb=18 155 592 560,clip]{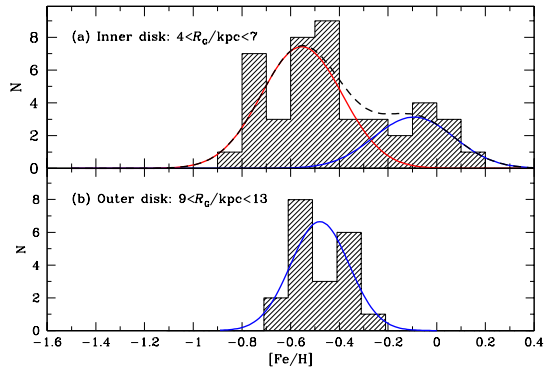}}
\caption{The metallicity distribution for the inner disk sample
show two Gaussians, one based on stars with $\rm [Mg/Fe]\geq0.2$
that has $\rm \langle[Fe/H]\rangle=-0.55\pm0.17$ (red curve), 
and one for stars with 
$\rm [Mg/Fe]\geq0.2$ that has $\rm \langle[Fe/H]\rangle=-0.09\pm0.17$
(blue curve). The Gaussian shown for the outer disk sample has 
$\rm \langle[Fe/H]\rangle=-0.48\pm0.12$.
\label{fig:mdf} 
}
\end{figure}

\subsection{Stellar parameters}

The 20 outer disk stars have effective temperatures and surface 
gravities in the ranges $3900 < \teff <5100$\,K and $0.9 < \log g < 3.2$, 
respectively, which is typical for K red giant stars, and similar
to the 44 inner disk giants from \cite{bensby2010letter}.

\subsection{Distances and kinematics}
\label{sec:distances}

The 19 of the 20 outer disk giants
have $R_{\rm G}$ between 11 and 13\,kpc, and one is very far away
at $R_{\rm G}\approx19$\,kpc (see Figs.~\ref{fig:glonglat}a
and \ref{fig:glonglat}b). Given the warp of the Galactic disk 
from \cite{momany2006}, and the scale-heights, scale-lengths,
and normalisations of the thin and thick disks in the solar neighbourhood 
from \cite{juric2008}, the dotted lines
in Fig.~\ref{fig:glonglat}a show the distance from the 
plane where the densities of thin and thick disk stars are 
equal.
At $X=5$\,kpc they are equal at 1.04\,kpc, at $X=8$\,kpc at 0.90\,kpc, 
and at $X=11$\,kpc at 0.75\,kpc.  Our inner and outer
disk samples have been observed without prior knowledge
of kinematics and/or metallicities. Therefore if the thin
and thick disks exist at these locations in the Galaxy and
follow the assumed scale-heights and scale-lengths, the 
stars in the inner and outer disk samples should consist of {\it both}
thin disk stars {\it and} thick disk stars. Actually,
Fig.~\ref{fig:glonglat}a indicates that
the outer disk sample should contain 13 thick disk stars and 7 thin 
disk stars, while the inner disk sample should contain 21 thick disk 
stars and 23 thin disk stars.

Based on kinematics, stars with 
total space velocities less than about $85\,\kms$ are generally associated with
the thin disk, while stars with higher velocities are
associated with the thick disk \citep[e.g.,][]{fuhrmann2004}.
These criteria are based on solar neighbourhood data, and if 
they are to be applied to stellar samples farther away, such as our inner and
outer disk giant samples, one assumes that  the properties of the
thin and thick disks in the solar neighbourhood are also valid there.
However, as showed by \cite{bensby2010rio}, it should be cautioned
that kinematical criteria can introduce significant mixing of the two 
populations as stars from the high-velocity tail of the thin disk are 
classified as thick disk stars (especially at high metallicities), 
and stars from the low-velocity tail 
of the thick disk as thin disk stars (espacially at low metallicities). 
Despite the large 
errors in the proper motions from \cite{zacharias2010}, which results
in very large errors in the space velocities, we show in
Figs.~\ref{fig:glonglat}c and \ref{fig:glonglat}d the Toomre
diagrams for the inner and outer disk stars, and it is clear that they
sample both the thin and the thick disk velocity spaces.

\section{Abundance results}
\label{sec:abundancetrends}

The abundance results for the outer disk giants are shown 
Fig.~\ref{fig:haltplottar}, where they are
compared to the \cite{bensby2010letter} inner disk red giant sample, 
and the \cite{alvesbrito2010} sample of
thin and thick disk red giants in the solar neighbourhood, and in Fig.~\ref{fig:mdf}
where we show the metallicity distribution functions. 
We stress again that all stars were analysed using exactly the same methods.

A first thing to notice is that the metallicity distributions functions 
(MDF) for the inner and outer disk samples are very different. The inner
disk MDF has a large spread ($\rm \langle[Fe/H]\rangle_{inner}=-0.42\pm 0.27$)
and suggests a bi-modal distribution, while the outer disk
MDF has a much smaller spread 
$\rm \langle[Fe/H]\rangle_{outer}=-0.48\pm 0.12$. 
Within the limited sample, the outer disk MDF is entirely consistent with 
a single value! The dispersion can be attributed solely to 
measurement uncertainties. 
Dividing the inner disk sample into two, one with stars that have
$\rm [Mg/Fe]\geq0.2$ (thick disk) and one with stars that have $\rm [Mg/Fe]<0.2$
(thin disk), results in two metallicity distributions with 
$\rm \langle[Fe/H]\rangle_{inner}=-0.55\pm 0.12$
and $\rm \langle[Fe/H]\rangle_{inner}=-0.09\pm 0.17$, respectively.

Regarding the outer disk, almost all stars have abundance ratios 
similar to what is seen in the nearby thin disk. This result is surprising 
because, based on the kinematics and the distances from the plane, 
a majority of the 20 stars should be thick disk 
stars. But only one, or maybe two, of the outer disk giants show thick 
disk abundance patterns. 

The abundance trends of the inner disk sample appears to contain 
stars with abundance patterns consistent with the nearby thin and thick disks. 
In \cite{bensby2010letter} we concluded that it is possible that a thin-thick
disk duality, similar to the one seen in the solar neighbourhood, is 
also present in the inner Galactic disk.

Figure~\ref{fig:scalelength}a shows the $\rm [\alpha/Fe]$ abundance 
ratio\footnote{$\alpha$ is here defined as the mean of Mg, Si and Ti.}
versus $R_{\rm G}$. All outer disk giants with 
distances beyond 11\,kpc have $\rm [\alpha/Fe]\approx 0.05$, while 
the inner disk giants have a larger spread and reaches 
$\rm [\alpha/Fe]\approx 0.4$. 
The one or two stars in the outer disk sample with elevated 
$\rm [\alpha/Fe]$ ratios are among the outer disk sample stars with smallest 
 $R_{\rm G}$. There appears to be a
sudden step in the $\rm [\alpha/Fe]$ abundance ratios at
$R_{\rm G}\approx 11$\,kpc, beyond which no stars with $\alpha$-enhancements
typical of the nearby thick disk stars can be seen. 

\section{Discussion}

The apparent lack of stars with thick disk chemistry in the outer
disk, even for stars high above the Galactic plane, yields
at least two possible interpretations:  A) The thick disk abundance
gradient is steeper than that of the thin disk, yielding degenerate
$\rm [\alpha/Fe]$ ratios around 12\,kpc; or more speculatively,
B) The thick disk's radial scale length is much shorter than that of 
the thin disk, so that most of our outer disk sample is dominated by 
thin disk stars.  In this section, we explore the second possibility.

\subsection{A short scale-length for the thick disk?}
\label{sec:scalelength}

\begin{figure}
\centering
\resizebox{\hsize}{!}{
\includegraphics[bb=18 165 592 400,clip]{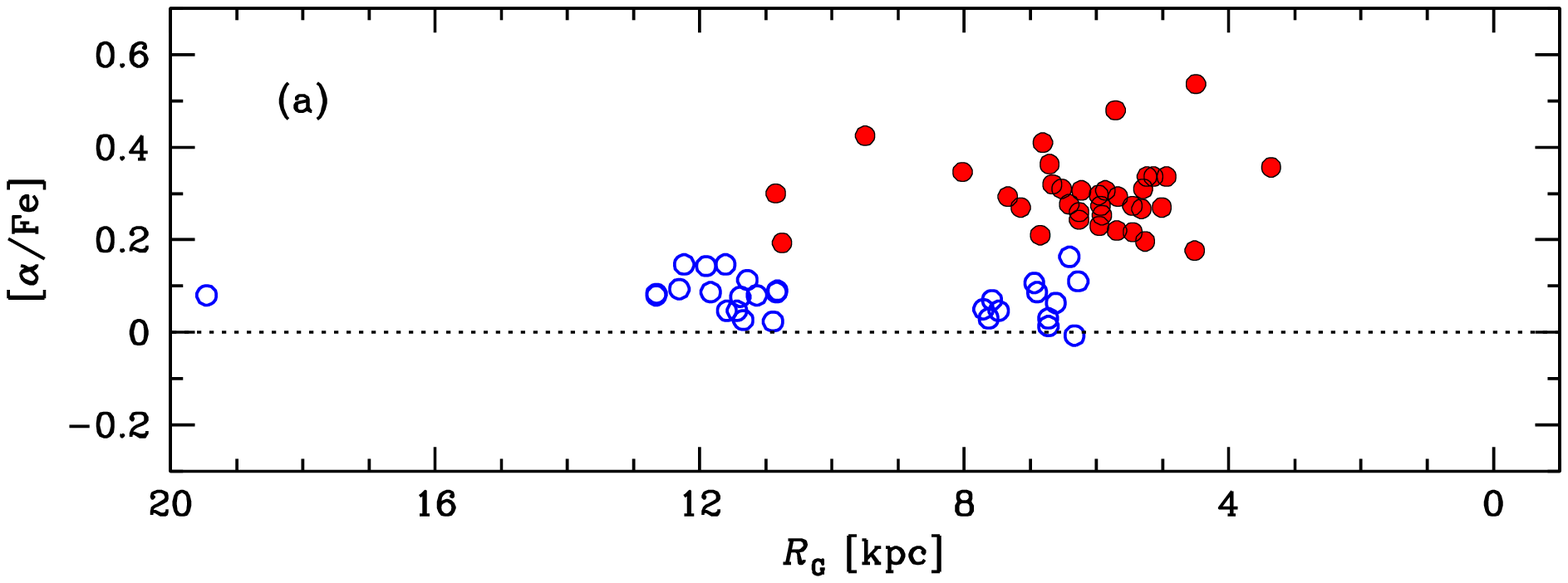}}
\resizebox{\hsize}{!}{
\includegraphics[bb=18 155 592 540,clip]{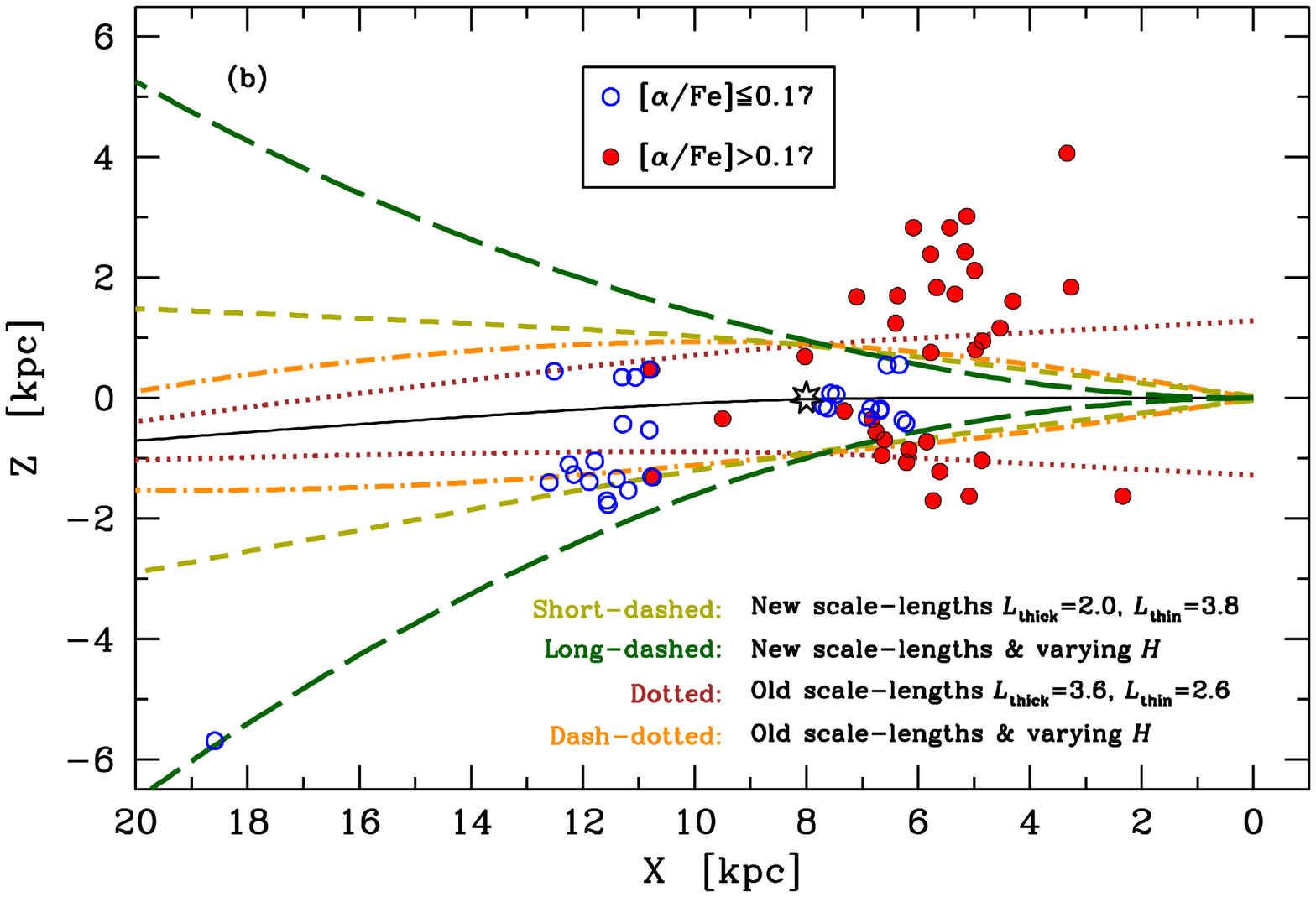}}
\caption{
(a) $\rm [\alpha/Fe]$ versus $R_{\rm G}$. 
(b) The inner and outer disk samples in the Galactic $X-Z$
coordinate system. The curves show loci where the two populations
have equal stellar density, for different assumed scale lengths $L$ and
scale heights $H$ (see Sect.~\ref{sec:scalelength} for further
details). In both (a) and (b) stars
with $\rm [\alpha/Fe]>0.17$ are marked by filled red circles, while stars
with  $\rm [\alpha/Fe]\leq0.17$ are marked by open blue circles.
\label{fig:scalelength} 
}
\end{figure}

A shorter scale-length for the thick disk means that the
thick disk will be more dominant in the inner disk, and the thin disk
will be more dominant in the outer disk. 
This is illustrated by the short-dashed line in 
Fig.~\ref{fig:scalelength}b where the thick disk scale-length has been 
changed so that a majority of the outer disk stars are within the
limits where the thick disk stars starts to dominate. \cite{juric2008} found 
that the scale-lengths for the thin and thick disks were anti-correlated, so 
when decreasing the thick disk scale-length  from 3.6\,kpc to 2.0\,kpc, 
we simultaneously increased 
thin disk scale-length from 2.6\,kpc to 3.8\,kpc.

In Fig.~\ref{fig:scalelength}b stars with $\rm [\alpha/Fe]>0.17$ are marked
by red solid circles and stars with 
$\rm [\alpha/Fe]\leq0.17$ by open blue circles. With the new scale-lengths,
we see that also for the inner disk
sample, the new scale-lengths appear to better match chemistry vs. vertical
distance from the Galactic plane.

So far, we have kept the vertical scale-heights fixed for the two disks,
assuming that they do not vary with $R_{\rm G}$. This is most likely not
the case. For instance, in a minor 
merger formation scenario for the thick disk, 
\cite{qu2011} shows that the thick disk scale-height increases linearly
with  $R_{\rm G}/L$. The orange dash-dotted lines in 
Fig.~\ref{fig:scalelength}b show how the equal density 
loci changes when keeping the old scale-lengths from \cite{juric2008}
{\it and} varying the scale-heights linearly with $R_{\rm G}/L$. The vertical 
distance from the plane where the thin disk dominates 
will only slightly increase with $R_{\rm G}$, and further out it will 
decrease again. Also shown in Fig.~\ref{fig:scalelength}b is
the case when adopting the {\it new} scale-lengths {\it and} varying 
the scale-heights with $R_{\rm G}$ (green long-dashed lines). 
The vertical distances from 
the plane where the thin disk
dominates will now increase even more than when keeping the scale-heights
fixed. Even the very distant giant star at $R_{\rm G}\approx 20$\,kpc is now
within the thin disk dominated region.
In all cases a constant has been multiplied to the scale-height functions
so that the normalisations and scale-lengths in the solar neighbourhood 
are reproduced.

\begin{table*}
\scriptsize
\centering
\caption{
Stellar parameters, kinematics, and abundances.
\label{tab:parameters}
}
\setlength{\tabcolsep}{1mm}
\begin{tabular}{ccccccccccccccc}
\tableline
\tableline
\noalign{\smallskip}
  Object     &
  $l$        &
  $b$        &
  $d$        &
  $V_{r}$    &
  $\ulsr$    &
  $\vlsr$    &
  $\wlsr$    &
$\teff$ &
$\log g$   &
[Fe/H]     &
[Mg/Fe]    &
[Si/Fe]    &
[Ti/Fe]   \\    
\noalign{\smallskip}
\tableline
\noalign{\smallskip}
  04342992$+$0306013 & 192.7 & $-$28.4 & 3.6 & 40.8 & $-$21 & $-$31 & $-$4 &  4100    &  1.2    &  $-$0.52  &  0.11   & 0.01  &  0.02 \\
  14053981$+$1304554 & 329.4 &  45.9   & 2.3 & 18.2 & $-$82 & $-$79 &   65 &  4200    &  1.9    &  $-$0.45  &  0.25   & 0.27  &  0.29 \\
\noalign{\smallskip}
\tableline
\end{tabular}
\tablecomments{Table 1 is published in its entirety in the electronic edition of the Astrophysical Journal.
A portion is shown here for guidance regarding its form and content.
The table
also includes data for the 44 inner disk giants from \cite{bensby2010letter}}
\end{table*}

\subsection{Metallicity gradients in the disks}
\label{sec:gradient}

The average metallicity of the thin disk in the solar neighbourhood
is $\rm \langle [Fe/H]\rangle = -0.06\pm 0.22$ \citep{casagrande2011}.
Assuming that the outer disk sample is purely thin disk, the average metallicity
of the thin disk at $R_{\rm G}\approx 11$\,kpc is 
$\rm \langle [Fe/H]\rangle = -0.48\pm 0.12$. This implies that there
is a strong abundance gradient in the thin disk, going from $R_{\rm G}=8$\,kpc
to $R_{\rm G}=11$\,kpc, fully consistent with
other studies of young stellar tracers \citep[see, e.g.,][]{cescutti2007}.

The solar neighbourhood thick disk MDF peaks at $\rm [Fe/H]=-0.60$ 
\citep[e.g.,][]{lee2011}. The metal-poor part of our bi-modal inner disk 
is likely associated with the thick disk
(see Fig.~\ref{fig:haltplottar}) and has an average metallicity
of $\rm [Fe/H]=-0.55$. These inner
thick disk stars are on average located at $R_{\rm G}\approx6$\,kpc. 
Over this 2\,kpc radial baseline, the thick disk seems to lack an 
abundance/metallicity gradient all together. 

Looking further into the inner regions of the Galaxy, recent studies of 
microlensed dwarf stars in the bulge has
shown that the bulge MDF is bi-modal with one peak at $\rm [Fe/H]\approx-0.6$
and another at $\rm [Fe/H]\approx+0.3$. In-between the peaks there is a void 
where no microlensed dwarf stars have been found \citep{bensby2010,bensby2011}.
The stars in the metal-poor bulge shows the same abundance trends, same average 
metallicities, and have the same age structure as the thick disk. 
These similarities have been seen in studies of giants as well
\citep[e.g.,][]{alvesbrito2010}. Although currently it is only possible
to state that the thick disk and metal-poor bulge has experienced similar
chemical histories,  this intriguing similarity may suggest that they are indeed the 
same population.
As the metal-poor bulge population peaks at $\rm \langle[Fe/H]\rangle=-0.60$ 
\citep{bensby2011}, which is identical to the value for the thick disk
in the solar neighbourhood, this would mean that the thick disk within the solar 
radius, all the way into the Galactic centre, is completely homogeneous. 
No metallicity, abundance, or age gradients whatsoever!

Could it be possible that radial migration is the cause for the
lack of gradients in the thick disk? 
Radial migration describes processes that cause the orbit of a 
star to change such that information on their formation radius is lost
and was first discussed by \cite{sellwood2002}.
And certainly, radial migration has surfaced as an 
important mechanism in the modelling of disk galaxies 
\citep[e.g.,][]{roskar2008b,roskar2008a,schonrich2009a,schonrich2009b,loebman2010}.
As radial migration of stellar orbits is a relatively
slow process, it should have affected old stellar populations more
than young populations. This might suggest that the metal-poor bulge
and the thick disk were formed at the same time and that the thick disk's 
radial metallicity gradient (if any was present) has been washed out over time, 
while in the younger thin disk the gradient is still present.

We notice that the probability of radial migration steeply decreases 
with increasing radial distance \citep[e.g.,][Fig.~4, left]{bird2011} for 
an isolated galaxy. However, mergers can make the migration probability 
much flatter, so a wide variety of results should be possible just 
by introducing mergers as an additional parameter. Anyway, assuming 
no significant mergers then migration should be much more efficient 
in the inner Galaxy, and the contribution from the inner disk will 
become progressively less and less important for the outer disk. 
We further note that \cite{lepine2011} emphasise that it is rather
easy to mix material from the inner regions up to about the solar distance, 
but that at around 8.5\,kpc from the Galactic centre there is a circular 
region void of material, the co-rotation gap \citep[e.g.,][]{amores2009}, 
and that mixing between the inner ($R_{\rm G} < 8.5$\,kpc) and outer 
($R_{\rm G} > 9$\,kpc) parts would be more difficult.

Furthermore, studies of radial colour and stellar mass density profiles for
external late-type spiral galaxies have revealed that as many as 90\,\%
have light profiles that can be classified as broken exponentials
\citep[e.g.,][]{bakos2008}. Their interpretation is that the break is due
to a radial change in the stellar population, rather than being a drop in the
distribution of mass. Also, resolved population studies from HST
show radial breaks in the stellar populations \citep[e.g.,][]{dejong2007}. 
This might be similar to what we see in
the Milky Way, an inner disk region (coupled to the bulge/thick disk
stellar populations) which has been more affected
by radial migration than the thin disk stellar population.

\section{Conclusions}

We have presented a detailed elemental abundance analysis of 20 giants
in the outer Galactic disk. Our results unambiguously show a lack
of stars with thick disk chemical patterns in the outer disk, 
even for stars very far from the Galactic plane. While this does not necessarily
imply anything about the structure of the two disk populations, 
we propose that this reflects a major 
difference in the scale-lengths between these components, and that the 
thick disk scale-length is significantly shorter than that of the thin disk.
We make a first estimate and find that a scale-length of $L_{thick}=2.0$\,kpc
for the thick disk and $L_{thin}=3.8$\,kpc for the thin disk are good
matches to our observations.

There is increasing evidence of a connection between the metal-poor
bulge and the thick disk, regarding their abundance patterns, 
metallicity distribution functions, ages and a flat radial
abundance gradient in the thick disk 
\citep[][this work]{melendez2008,alvesbrito2010,bensby2010,bensby2011}. 
Stellar radial migration could plausibly explain the lack of 
radial gradients in the thick disk over the Galaxy's
history. All evidence indicates
a link between both populations, pointing to a shared origin.

\acknowledgements
T.B. was funded by grant No. 621-2009-3911 
from The Swedish Research Council. This work was also supported by 
the NSF, grant AST-0448900 to MSO. AAB acknowledges grants from
FONDECYT (process 3100013). J.M. thanks support from FAPESP (2010/50930-6),
USP (Novos Docentes) and CNPq (Bolsa de produtividade). We also thank
Victor Debattista for comments on a draft 
version of the paper.



\begin{thebibliography}{46}
\expandafter\ifx\csname natexlab\endcsname\relax\def\natexlab#1{#1}\fi

\bibitem[{{Alves-Brito} {et~al.}(2010){Alves-Brito}, {Mel{\'e}ndez}, {Asplund},
  {Ram{\'{\i}}rez}, \& {Yong}}]{alvesbrito2010}
{Alves-Brito}, A., {Mel{\'e}ndez}, J., {Asplund}, M., {Ram{\'{\i}}rez}, I., \&
  {Yong}, D. 2010, \aap, 513, A35

\bibitem[{{Am{\^o}res} {et~al.}(2009){Am{\^o}res}, {L{\'e}pine}, \&
  {Mishurov}}]{amores2009}
{Am{\^o}res}, E.~B., {L{\'e}pine}, J.~R.~D., \& {Mishurov}, Y.~N. 2009, \mnras,
  400, 1768

\bibitem[{{Andrievsky} {et~al.}(2004){Andrievsky}, {Luck}, {Martin}, \&
  {L{\'e}pine}}]{andrievsky2004}
{Andrievsky}, S.~M., {Luck}, R.~E., {Martin}, P., \& {L{\'e}pine}, J.~R.~D.
  2004, \aap, 413, 159

\bibitem[{{Bakos} {et~al.}(2008){Bakos}, {Trujillo}, \& {Pohlen}}]{bakos2008}
{Bakos}, J., {Trujillo}, I., \& {Pohlen}, M. 2008, \apjl, 683, L103

\bibitem[{{Bensby} {et~al.}(2011){Bensby}, {Ad{\'e}n}, {Mel\'endez}, {Gould},
  {Feltzing}, {Asplund}, {Johnson}, {Lucatello}, {Yee}, {Ram\'irez}, {Cohen},
  {Gal-Yam}, \& {Bond}}]{bensby2011}
{Bensby}, T., {Ad{\'e}n}, D., {Mel\'endez}, J., {Gould}, A., {Feltzing}, S.,
  {Asplund}, M., {Johnson}, J.~A., {Lucatello}, S., {Yee}, J.~C., {Ram\'irez},
  I., {Cohen}, J.~{Thompson}, I., {Gal-Yam}, A.~{Sumi}, T., \& {Bond}, I. 2011,
  \aap, submitted

\bibitem[{{Bensby} {et~al.}(2010{\natexlab{a}}){Bensby}, {Alves-Brito}, {Oey},
  {Yong}, \& {Mel{\'e}ndez}}]{bensby2010letter}
{Bensby}, T., {Alves-Brito}, A., {Oey}, M.~S., {Yong}, D., \& {Mel{\'e}ndez},
  J. 2010{\natexlab{a}}, \aap, 516, L13

\bibitem[{{Bensby} \& {Feltzing}(2010)}]{bensby2010rio}
{Bensby}, T., \& {Feltzing}, S. 2010, in IAU Symposium, Vol. 265, IAU
  Symposium, ed. {K.~Cunha, M.~Spite, \& B.~Barbuy}, 300--303

\bibitem[{{Bensby} {et~al.}(2010{\natexlab{b}}){Bensby}, {Feltzing}, {Johnson},
  {Gould}, {Ad{\'e}n}, M., {Mel\'endez}, {Gal-Yam}, {Lucatello}, {Sana},
  {Sumi}, {Miyake}, {Suzuki}, {Han}, {Bond}, \& {Udalski}}]{bensby2010}
{Bensby}, T., {Feltzing}, S., {Johnson}, J.~A., {Gould}, A., {Ad{\'e}n}, D.,
  M., A., {Mel\'endez}, J., {Gal-Yam}, A., {Lucatello}, S., {Sana}, H., {Sumi},
  T., {Miyake}, N., {Suzuki}, D., {Han}, C., {Bond}, I., \& {Udalski}, A.
  2010{\natexlab{b}}, \aap, 512, A41

\bibitem[{{Bensby} {et~al.}(2003){Bensby}, {Feltzing}, \& {Lundstr{\"
  o}m}}]{bensby2003}
{Bensby}, T., {Feltzing}, S., \& {Lundstr{\" o}m}, I. 2003, \aap, 410, 527

\bibitem[{{Bensby} {et~al.}(2004){Bensby}, {Feltzing}, \& {Lundstr{\"
  o}m}}]{bensby2004}
---. 2004, \aap, 415, 155

\bibitem[{{Bensby} {et~al.}(2005){Bensby}, {Feltzing}, {Lundstr{\" o}m}, \&
  {Ilyin}}]{bensby2005}
{Bensby}, T., {Feltzing}, S., {Lundstr{\" o}m}, I., \& {Ilyin}, I. 2005, \aap,
  433, 185

\bibitem[{{Bensby} {et~al.}(2007){Bensby}, {Zenn}, {Oey}, \&
  {Feltzing}}]{bensby2007letter2}
{Bensby}, T., {Zenn}, A.~R., {Oey}, M.~S., \& {Feltzing}, S. 2007, \apjl, 663,
  L13

\bibitem[{{Bernstein} {et~al.}(2003){Bernstein}, {Shectman}, {Gunnels},
  {Mochnacki}, \& {Athey}}]{bernstein2003}
{Bernstein}, R., {Shectman}, S.~A., {Gunnels}, S.~M., {Mochnacki}, S., \&
  {Athey}, A.~E. 2003, in Proceedings of the SPIE, Volume 4841, ed. M.~{Iye} \&
  A.~F.~M. {Moorwood}, 1694--1704

\bibitem[{{Bird} {et~al.}(2011){Bird}, {Kazantzidis}, \& {Weinberg}}]{bird2011}
{Bird}, J.~C., {Kazantzidis}, S., \& {Weinberg}, D.~H. 2011, ArXiv e-prints

\bibitem[{{Burstein}(1979)}]{burstein1979}
{Burstein}, D. 1979, \apj, 234, 829

\bibitem[{{Carney} {et~al.}(2005){Carney}, {Yong}, {Teixera de Almeida}, \&
  {Seitzer}}]{carney2005}
{Carney}, B.~W., {Yong}, D., {Teixera de Almeida}, M.~L., \& {Seitzer}, P.
  2005, \aj, 130, 1111

\bibitem[{{Carraro} {et~al.}(2007){Carraro}, {Geisler}, {Villanova},
  {Frinchaboy}, \& {Majewski}}]{carraro2007}
{Carraro}, G., {Geisler}, D., {Villanova}, S., {Frinchaboy}, P.~M., \&
  {Majewski}, S.~R. 2007, \aap, 476, 217

\bibitem[{{Casagrande} {et~al.}(2011){Casagrande}, {Sch{\"o}nrich}, {Asplund},
  {Cassisi}, {Ramirez}, {Melendez}, {Bensby}, \& {Feltzing}}]{casagrande2011}
{Casagrande}, L., {Sch{\"o}nrich}, R., {Asplund}, M., {Cassisi}, S., {Ramirez},
  I., {Melendez}, J., {Bensby}, T., \& {Feltzing}, S. 2011, \aap, 530, A138

\bibitem[{{Cescutti} {et~al.}(2007){Cescutti}, {Matteucci}, {Fran{\c c}ois}, \&
  {Chiappini}}]{cescutti2007}
{Cescutti}, G., {Matteucci}, F., {Fran{\c c}ois}, P., \& {Chiappini}, C. 2007,
  \aap, 462, 943

\bibitem[{{Daflon} \& {Cunha}(2004)}]{daflon2004}
{Daflon}, S., \& {Cunha}, K. 2004, \apj, 617, 1115

\bibitem[{{Daflon} {et~al.}(2004){Daflon}, {Cunha}, \&
  {Butler}}]{daflon2004outer}
{Daflon}, S., {Cunha}, K., \& {Butler}, K. 2004, \apj, 606, 514

\bibitem[{{de Jong} {et~al.}(2007){de Jong}, {Seth}, {Radburn-Smith}, {Bell},
  {Brown}, {Bullock}, {Courteau}, {Dalcanton}, {Ferguson}, {Goudfrooij},
  {Holfeltz}, {Holwerda}, {Purcell}, {Sick}, \& {Zucker}}]{dejong2007}
{de Jong}, R.~S., {Seth}, A.~C., {Radburn-Smith}, D.~J., {Bell}, E.~F.,
  {Brown}, T.~M., {Bullock}, J.~S., {Courteau}, S., {Dalcanton}, J.~J.,
  {Ferguson}, H.~C., {Goudfrooij}, P., {Holfeltz}, S., {Holwerda}, B.~W.,
  {Purcell}, C., {Sick}, J., \& {Zucker}, D.~B. 2007, \apjl, 667, L49

\bibitem[{{Fuhrmann}(2004)}]{fuhrmann2004}
{Fuhrmann}, K. 2004, Astronomische Nachrichten, 325, 3

\bibitem[{{Fuhrmann}(2008)}]{fuhrmann2008}
---. 2008, \mnras, 384, 173

\bibitem[{{Gilmore} \& {Reid}(1983)}]{gilmore1983}
{Gilmore}, G., \& {Reid}, N. 1983, \mnras, 202, 1025

\bibitem[{{Gilmore} {et~al.}(1995){Gilmore}, {Wyse}, \& {Jones}}]{gilmore1995}
{Gilmore}, G., {Wyse}, R.~F.~G., \& {Jones}, J.~B. 1995, \aj, 109, 1095

\bibitem[{{Jacobson} {et~al.}(2011){Jacobson}, {Friel}, \&
  {Pilachowski}}]{jacobson2011}
{Jacobson}, H.~R., {Friel}, E.~D., \& {Pilachowski}, C.~A. 2011, \aj, 141, 58

\bibitem[{{Juri{\'c}} {et~al.}(2008){Juri{\'c}}, {Ivezi{\'c}}, {Brooks},
  {Lupton}, {Schlegel}, {Finkbeiner}, {Padmanabhan}, {Bond}, {Sesar},
  {Rockosi}, {Knapp}, {Gunn}, {Sumi}, {Schneider}, {Barentine}, {Brewington},
  {Brinkmann}, {Fukugita}, {Harvanek}, {Kleinman}, {Krzesinski}, {Long},
  {Neilsen}, {Nitta}, {Snedden}, \& {York}}]{juric2008}
{Juri{\'c}}, M., {Ivezi{\'c}}, {\v Z}., {Brooks}, A., {Lupton}, R.~H.,
  {Schlegel}, D., {Finkbeiner}, D., {Padmanabhan}, N., {Bond}, N., {Sesar}, B.,
  {Rockosi}, C.~M., {Knapp}, G.~R., {Gunn}, J.~E., {Sumi}, T., {Schneider},
  D.~P., {Barentine}, J.~C., {Brewington}, H.~J., {Brinkmann}, J., {Fukugita},
  M., {Harvanek}, M., {Kleinman}, S.~J., {Krzesinski}, J., {Long}, D.,
  {Neilsen}, Jr., E.~H., {Nitta}, A., {Snedden}, S.~A., \& {York}, D.~G. 2008,
  \apj, 673, 864

\bibitem[{{Lee} {et~al.}(2011){Lee}, {Beers}, {An}, {Ivezic}, {Just},
  {Rockosi}, {Morrison}, {Johnson}, {Schonrich}, {Bird}, {Yanny}, {Harding}, \&
  {Rocha-Pinto}}]{lee2011}
{Lee}, Y.~S., {Beers}, T.~C., {An}, D., {Ivezic}, Z., {Just}, A., {Rockosi},
  C.~M., {Morrison}, H.~L., {Johnson}, J.~A., {Schonrich}, R., {Bird}, J.,
  {Yanny}, B., {Harding}, P., \& {Rocha-Pinto}, H.~J. 2011, ArXiv e-prints

\bibitem[{{L\'epine} {et~al.}(2011){L\'epine}, {Cruz}, {Scarano Jr}, {Barros},
  {Dias}, {Pomp\'eia}, {Andrievsky}, {Carraro}, \& {Famaey}}]{lepine2011}
{L\'epine}, J.~R.~D., {Cruz}, P., {Scarano Jr}, S., {Barros}, D.~A., {Dias},
  W.~S., {Pomp\'eia}, L., {Andrievsky}, S.~M., {Carraro}, G., \& {Famaey}, B.
  2011, \mnras, submitted

\bibitem[{{Loebman} {et~al.}(2010){Loebman}, {Roskar}, {Debattista}, {Ivezic},
  {Quinn}, \& {Wadsley}}]{loebman2010}
{Loebman}, S.~R., {Roskar}, R., {Debattista}, V.~P., {Ivezic}, Z., {Quinn},
  T.~R., \& {Wadsley}, J. 2010, arXiv:1009.5997 [astro-ph.GA]

\bibitem[{{Luck} {et~al.}(2006){Luck}, {Kovtyukh}, \& {Andrievsky}}]{luck2006}
{Luck}, R.~E., {Kovtyukh}, V.~V., \& {Andrievsky}, S.~M. 2006, \aj, 132, 902

\bibitem[{{Mel{\'e}ndez} {et~al.}(2008){Mel{\'e}ndez}, {Asplund},
  {Alves-Brito}, {Cunha}, {Barbuy}, {Bessell}, {Chiappini}, {Freeman},
  {Ram{\'{\i}}rez}, {Smith}, \& {Yong}}]{melendez2008}
{Mel{\'e}ndez}, J., {Asplund}, M., {Alves-Brito}, A., {Cunha}, K., {Barbuy},
  B., {Bessell}, M.~S., {Chiappini}, C., {Freeman}, K.~C., {Ram{\'{\i}}rez},
  I., {Smith}, V.~V., \& {Yong}, D. 2008, \aap, 484, L21

\bibitem[{{Momany} {et~al.}(2006){Momany}, {Zaggia}, {Gilmore}, {Piotto},
  {Carraro}, {Bedin}, \& {de Angeli}}]{momany2006}
{Momany}, Y., {Zaggia}, S., {Gilmore}, G., {Piotto}, G., {Carraro}, G.,
  {Bedin}, L.~R., \& {de Angeli}, F. 2006, \aap, 451, 515

\bibitem[{{Qu} {et~al.}(2011){Qu}, {Di Matteo}, {Lehnert}, \& {van
  Driel}}]{qu2011}
{Qu}, Y., {Di Matteo}, P., {Lehnert}, M.~D., \& {van Driel}, W. 2011, \aap,
  530, A10+

\bibitem[{{Reddy} {et~al.}(2006){Reddy}, {Lambert}, \& {Allende
  Prieto}}]{reddy2006}
{Reddy}, B.~E., {Lambert}, D.~L., \& {Allende Prieto}, C. 2006, \mnras, 367,
  1329

\bibitem[{{Ro{\v s}kar} {et~al.}(2008{\natexlab{a}}){Ro{\v s}kar},
  {Debattista}, {Quinn}, {Stinson}, \& {Wadsley}}]{roskar2008b}
{Ro{\v s}kar}, R., {Debattista}, V.~P., {Quinn}, T.~R., {Stinson}, G.~S., \&
  {Wadsley}, J. 2008{\natexlab{a}}, \apjl, 684, L79

\bibitem[{{Ro{\v s}kar} {et~al.}(2008{\natexlab{b}}){Ro{\v s}kar},
  {Debattista}, {Stinson}, {Quinn}, {Kaufmann}, \& {Wadsley}}]{roskar2008a}
{Ro{\v s}kar}, R., {Debattista}, V.~P., {Stinson}, G.~S., {Quinn}, T.~R.,
  {Kaufmann}, T., \& {Wadsley}, J. 2008{\natexlab{b}}, \apjl, 675, L65

\bibitem[{{Sch{\"o}nrich} \& {Binney}(2009{\natexlab{a}})}]{schonrich2009a}
{Sch{\"o}nrich}, R., \& {Binney}, J. 2009{\natexlab{a}}, \mnras, 396, 203

\bibitem[{{Sch{\"o}nrich} \& {Binney}(2009{\natexlab{b}})}]{schonrich2009b}
---. 2009{\natexlab{b}}, \mnras, 399, 1145

\bibitem[{{Sellwood} \& {Binney}(2002)}]{sellwood2002}
{Sellwood}, J.~A., \& {Binney}, J.~J. 2002, \mnras, 336, 785

\bibitem[{{Twarog} {et~al.}(1997){Twarog}, {Ashman}, \&
  {Anthony-Twarog}}]{twarog1997}
{Twarog}, B.~A., {Ashman}, K.~M., \& {Anthony-Twarog}, B.~J. 1997, \aj, 114,
  2556

\bibitem[{{Wyse} \& {Gilmore}(1995)}]{wyse1995}
{Wyse}, R.~F.~G., \& {Gilmore}, G. 1995, \aj, 110, 2771

\bibitem[{{Yong} {et~al.}(2005){Yong}, {Carney}, \& {Teixera de
  Almeida}}]{yong2005}
{Yong}, D., {Carney}, B.~W., \& {Teixera de Almeida}, M.~L. 2005, \aj, 130, 597

\bibitem[{{Yong} {et~al.}(2006){Yong}, {Carney}, {Teixera de Almeida}, \&
  {Pohl}}]{yong2006}
{Yong}, D., {Carney}, B.~W., {Teixera de Almeida}, M.~L., \& {Pohl}, B.~L.
  2006, \aj, 131, 2256

\bibitem[{{Zacharias} {et~al.}(2010){Zacharias}, {Finch}, {Girard}, {Hambly},
  {Wycoff}, {Zacharias}, {Castillo}, {Corbin}, {DiVittorio}, {Dutta}, {Gaume},
  {Gauss}, {Germain}, {Hall}, {Hartkopf}, {Hsu}, {Holdenried}, {Makarov},
  {Martines}, {Mason}, {Monet}, {Rafferty}, {Rhodes}, {Siemers}, {Smith},
  {Tilleman}, {Urban}, {Wieder}, {Winter}, \& {Young}}]{zacharias2010}
{Zacharias}, N., {Finch}, C., {Girard}, T., {Hambly}, N., {Wycoff}, G.,
  {Zacharias}, M.~I., {Castillo}, D., {Corbin}, T., {DiVittorio}, M., {Dutta},
  S., {Gaume}, R., {Gauss}, S., {Germain}, M., {Hall}, D., {Hartkopf}, W.,
  {Hsu}, D., {Holdenried}, E., {Makarov}, V., {Martines}, M., {Mason}, B.,
  {Monet}, D., {Rafferty}, T., {Rhodes}, A., {Siemers}, T., {Smith}, D.,
  {Tilleman}, T., {Urban}, S., {Wieder}, G., {Winter}, L., \& {Young}, A. 2010,
  \aj, 139, 2184

\end{thebibliography}
\end{document}